\documentclass[aps,prl,10pt,twocolumn,showpacs,superscriptaddress]{revtex4-1}
\usepackage{amsmath}
\usepackage{latexsym}
\usepackage{amssymb}
\usepackage{bm}
\usepackage{graphics,epstopdf}
\usepackage{color}
\usepackage{hyperref}

\usepackage{newlfont}
\usepackage{amsfonts}
\usepackage{amsthm}
\usepackage{graphicx}
\usepackage{epsfig}

\newcommand{\ket}[1]{|{#1}\rangle}

\newcommand{\expec}[1]{\langle#1\rangle}

\usepackage{times}

\usepackage[up]{subfigure}

\newcommand{\be}{\begin{equation}}
\newcommand{\ee}{\end{equation}}
\newcommand{\bc}{\begin{center}}
\newcommand{\ec}{\end{center}}
\newcommand{\bea}{\begin{eqnarray}}
\newcommand{\eea}{\end{eqnarray}}
\newcommand{\ba}{\begin{array}}
\newcommand{\ea}{\end{array}}

\begin{document}
\title{Large Polarization Squeezing in Non-Degenerate Parametric \\ Amplification of Coherent Radiation}

\author{Namrata Shukla}
\email{namratashukla@hri.res.in}

\affiliation{Now at, Quantum Information and Computation Group,\\
Harish-Chandra Research Institute, Chhatnag Road, Jhunsi,
Allahabad 211 019, India}

\affiliation{Department of Physics,\\
University of Allahabad, Allahabad, Allahabad-211001, UP, India}

\author{Ranjana Prakash}


\affiliation{Department of Physics,\\
University of Allahabad, Allahabad, Allahabad-211001, UP, India}



\date{\today}

\begin{abstract}
Polarization squeezing is shown to occur in non-degenerate
parametric amplification of coherent light and the degree of squeezing
at interaction time $T$ can be as large as $1-e^{-2T}$. This gives $86.4\%$ polarization squeezing 
for $T=1$ and $98.2\%$ for $T=2$. One simple case when this occurs is on taking initially plane polarized 
light having equal amplitudes in the two modes that finally has equal intensities of two circular 
polarizations. This suggest the experimental settings of parameters to achieve this extent of
polarization squeezing in coherent light. 
\end{abstract}

\pacs{Valid PACS appear here}
\maketitle

\section*{INTRODUCTION}
Polarization of light is a concept receiving attention since very long 
and the mathematically convenient way of describing partial polarization 
was given classically with the help of Stokes parameters \cite{1,2}. For a monochromatic 
unidirectional light traveling along z-direction, Stokes parameters denoted by
$S_{0}$ and ${\bm S}=S_{1}, S_{2}, S_{3}$ are defined as
\begin{equation}
\label{eq1}
S_{0,1}= \expec{\bm{\mathcal{E}}_{x}^{*}\bm{\mathcal{E}}_{x}}
\pm\expec{\bm{\mathcal{E}}_{y}^{*}\bm{\mathcal{E}}_{y}},~
S_{2}+i S_{3}=2 \expec{\bm{\mathcal{E}}_{x}^{*}\bm{\mathcal{E}}_{y}},
\end{equation}
where $ {\bm{\mathcal E}_{x, y}}$ are the components of analytic signal \cite{3} 
for the electric field. For perfectly polarized light
\begin{equation}
\label{eq2}
 S_{0}^2=|{\bm S}|^2=S_{1}^2+S_{2}^2+S_{3}^2.
\end{equation}
The point having coordinates $(S_1, S_2, S_3) $ lies on a sphere of radius $S_{0}$ called Poincare sphere.
For unpolarized light, $ |{\bm S}|=0 $ and for partially polarized light,
 $ S_0>|{\bm S}| $.

Relevant continuous variables for the treatment of quantum nature 
of polarization of the system are called quantum Stokes operators \cite{4}. 
Quantum Stokes operators and the associated Poincare sphere 
describes the quantum nature of polarization of light. In direct analogy with the 
classical Stokes parameters, these Stokes operators $ \hat S_0 $ and
$\hat{\bm S}=\hat S_{1,2, 3}$ are defined in terms of creation and 
annihilation operators as
\begin {equation}
\label{eq3}
\hat S_{0, 1}=\hat a_{x}^\dagger \hat a_{x}\pm \hat a_{y}^\dagger \hat a_{y},~
\hat S_{2}+i \hat S_{3}=2 \hat a_{x}^\dagger \hat a_{y}.
\end {equation}
These Stokes operators involve the coherence functions \cite{5} of order $(1,1)$
and shown to be insufficient to describe polarization completely as $ \bm S=0 $ does 
not represent only the unpolarized light \cite{6}. These operators however, remain important 
because of the non-classicalities like polarization squeezing and 
polarization entanglement associated with polarization. 

Commutation relations $ [\hat a_{j}, \hat a_{k}^\dagger]=\delta_{jk} $ of annihilation 
and creation operators lead to the commutation relations
\begin{equation}
\label{eq4}
[\hat S_0, \hat S_j]=0,~[\hat S_j, \hat S_k]=2i{\sum_{l}\epsilon_{jkl}}~\hat S_{l},
\end {equation}
$\epsilon_{jkl}$ being Levi-Civita symbol for $(j,k,l=1,2,$ or $3)$.
The quantum fluctuations of these Stokes operators are bounded below by following uncertainty relations
\begin{equation}
\label{eq5}
V_{j} V_{k}\geqslant{\expec{\hat S_{l}}}^2,~ 
V_{j}\equiv\expec{\hat S_{j}^2}- {\expec{\hat S_{j}}}^2.
\end{equation}

Squeezed radiation states in quantum optics are identified 
by the property that their quantum fluctuations are reduced 
below the standard quantum limit in one of the quadrature components. 
Polarization squeezing \cite{7} is defined using above mentioned commutation relations
and uncertainty products for Stokes operators in a similar pattern. Polarization squeezing 
is important in quantum information theory and it is desirable to devise methods for 
generation of states with appreciable amount of polarization squeezing. 

In a study by Heersink et al. \cite{8}, polarization squeezing has been investigated 
by taking both the polarization modes of coherent light to be amplitude squeezed. 
$\hat S_1 $ is found to be polarization squeezed by $-3.4 $ dB as per the definition of polarization 
squeezing given in the same paper and it was also experimentally investigated. 

In this paper, we investigate the polarization squeezing using the most general criterion, 
in the coherent light subjected to non-degenerate parametric amplification which is never reported 
before as a closed form result. Parametric amplification is used to generate high polarization squeezing 
which is more than earlier reported cases for considerably small times of interaction.The experimental
feasibility of the technique with suitable fixing of parameters is discussed in section IV.   

Coinciding with the above reported paper by Heersink et.al., this study reveals the polarization
squeezing only along $\hat S_1 $ component of Stokes vector. We quantify the polarization squeezing  
by defining Squeezing factor and Degree of squeezing. Minimum value of squeezing factor exhibits
the maximum squeezing as a function of interaction time.  


\section{Criterion for Polarization squeezing}
Polarization squeezing was first defined by Chirkin et al. \cite{9} in terms of
variances of stokes operators for a given state as
\begin {equation}
\label{eq6}
 V_{j}<V_{j} (coh)=\hat S_{0},
\end {equation}
,{\it i.e.}, $\hat S_{j}$ is squeezed if the variance of Stokes parameter is less than that for 
an equally intense coherent state.

Heersink et al. defined polarization squeezing in their earlier mentioned paper using the uncertainty relations
\eqref{eq5} in the form
\begin {equation}
\label{eq7}
 V_{j}<\mid\expec{\hat S_{l}}\mid<V_{k} ~~for~~ j\neq k \neq l,
\end {equation}
for squeezing of $ \hat S_{j}$. 

Luis and Korolkova \cite{10} then considered various criteria for polarization
squeezing and their stringency was compared. They give the following criterion for polarization squeezing
of a component of $\hat {\bm S}$ along a unit vector ${\bm n}$ as
\begin{equation}
\label{eq8}
 V_{\bm n}<\mid\expec{\hat S_{\bm n_{\perp}}}\mid,
\end{equation}
where $ \hat S_{\bm n_{\perp}}$ is component of $\hat {\bm S}$ along unit vector $\bm n_{\perp}$ 
which is perpendicular to $ \bm n $. For suitable orthogonal components $ \hat S_{\bm n} $ and 
$ \hat S_{\bm n_{\perp}}$, they have discussed the order of stringency of the various criteria
\begin{equation}
\label{eq9}
 V_{\bm n}<\expec{S_{\bm n_{\perp}}}^2 / \expec{\hat S_{0}}
 <\mid\expec{S_{\bm n_{\perp}}}\mid<\expec{\hat S_0}.
\end{equation}
The authors \cite{11} finally write the criterion for polarization squeezing in 
the form

\begin{eqnarray}
V_{\bm n}\equiv\expec{\Delta S_{\bm n}^2}&<{\mid\expec{S_{\bm n_{\perp}}}\mid}_{max} \\\nonumber
&=&\sqrt{{\mid\expec{\hat {\bm S}}\mid}^2-{\expec{S_{\bm n}}}^2} \label{eq10} 
\end{eqnarray}
arguing that for a given component $ \hat S_{\bm n} $ there are infinite directions 
$ \bm n_{\perp} $ and therefore it is required to consider the maximum possible value of 
$ \mid\expec{\hat S_{n_{\perp}}}\mid $.  
Eqs. \eqref{eq6} - \eqref{eq8} and \eqref{eq10} describes non-classicality but the uncertainty relations
are considered only in \eqref{eq7}, \eqref{eq8} and \eqref{eq10}. 

We shall use the criterion \eqref{eq10} for polarization squeezing which is most general and based 
on the actual uncertainty relations. We may define squeezing factor $ \mathcal{\bm S}_{\bm n} $ and 
degree of squeezing $ \mathcal{\bm D}_{\bm n} $ by writing
\begin{equation}
\label{eq11}
\mathcal{\bm S}_{\bm n}=\frac{V_{\bm n}}{\sqrt{{\mid\expec{\hat {\bm S}}\mid}^2-{\expec{S_{\bm n}}}^2}},~
\mathcal{\bm D}_{\bm n}=1-\mathcal{\bm S}_{\bm n}.
\end{equation}

Non-classicalities appear when $ 1>\mathcal{\bm S}_{\bm n}>0 $ and the degree of squeezing $ \mathcal{\bm D}_{\bm n}$
lies between $0$ and $1$.


\section{Hamiltonian and Polarization squeezing}

The hamiltonian \cite{12} for non-degenerate parametric amplification of light traveling along z-direction 
is given by
\begin{equation}
\label{eq12} H=k\big(\hat a_{x}^\dagger\hat a_{y}^\dagger+\hat a_{x}\hat a_{y}\big),
\end{equation}
$k$ being coupling constant and $ \hat a_{x,y} $ are annihilation operators for the two linear polarizations $x$ and $y$. 
Solution to this hamiltonian \cite{12} is given by
\begin{eqnarray}
\hat a_x(t)&=&(\cosh kt)\hat a_x-i(\sinh kt)\hat a_{y}^\dagger, \\\nonumber
\hat a_y(t)&=&(\cosh kt)\hat a_y-i(\sinh kt)\hat a_{x}^\dagger.\label{eq13}  
\end{eqnarray}
If we consider the incident light in the coherent state $ \ket{\alpha, \beta}$, where 
$ \alpha=A\cos{\theta}e^{i\phi_x}, \beta=A\sin{\theta}e^{i\phi_y}$, straight forward 
calculations give the expectation values and variances of Stokes parameters 
at the interaction time $ T\equiv kt $ as
\begin{eqnarray}
\expec{\hat S_1(T)}&=& A^2 \cos2\theta,\nonumber\\
\expec{\hat S_2(T)}&=& A^2[\cosh2kt \sin^2\theta \cos(\phi_x-\phi_y)\nonumber\\
&&-\sinh2kt(\cos^2\theta \sin2\phi_x \nonumber
+\sin^2\theta \sin2\phi_y)],\nonumber\\
\expec{\hat S_3(T)}&=&A^2[-\cosh2kt \sin^2\theta \sin(\phi_x-\phi_y)\nonumber\\
&&-\sinh2kt(\cos^2\theta \cos2\phi_x
-\sin^2\theta \cos2\phi_y)],~~~~\nonumber\\
\label{eq14}
\end{eqnarray}
and 
\begin{eqnarray}
V_1(T)&=&A^2,\nonumber\\
V_2(T)&=&V_3(T)\nonumber\\
&=&A^2 \cosh^22kt+\sinh^22kt(A^2+1) \nonumber\\
&&-A^2\sinh4kt(c^2+s^2)\sin2\theta \sin(\phi_x+\phi_y),~~~~~~~~\nonumber\\
\label{eq15}
\end{eqnarray}
where $c=\cosh kt,~s=\sinh kt$. \\

As per the criterion given by Eq. \eqref{eq10}, a glance at expressions for $ V_1 $ and $\expec{\hat S_{2,3}}$ gives an idea
that  component $\hat S_1 $ can be squeezed. The squeezing factor in this case takes the form
\begin{equation}
\mathcal{S}_{1}=\frac{A^2}{A^2\sqrt{R}},\nonumber
\end{equation}
where
\begin{eqnarray}
R&=&\frac{\sqrt{\expec{\hat S_2}^2+\expec{\hat S_3}^2}}{A^2}\nonumber\\
&=&\cosh^22kt\sin^22\theta+\sinh^22kt(\sin^4\theta+\cos^4\theta)\nonumber\\
&&-2\sinh^22kt\sin^2\theta\cos^2\theta\cos2(\phi_x+\phi_y)\nonumber\\
&&-\sinh4kt\sin2\theta\sin(\phi_x+\phi_y).\nonumber
\end{eqnarray}
In order to achieve maximum squeezing we need to minimize the squeezing factor $\mathcal{S}_{1}$ and 
thus maximize $R$. A slight manipulation lets $R$ be written as
\begin{equation}
\label{eq16}
R=[\cosh2T-\sinh2T \sin2\theta\sin(\phi_x+\phi_y)]^2-\cos^22\theta.
\end{equation}
This gives maximum value of $R$ and hence minimum value of squeezing factor $\mathcal{\bm S}_1 $
denoted by $\mathcal{S}_{1 min}$, for $\theta=\pi/4$ and $(\phi_x+\phi_y)=3\pi/2$ in their admissible range of values. It leads to, 
\begin{equation}
\label{eq17}
\mathcal{S}_{1 min}=e^{-2T}.
\end{equation}
\begin{figure}[th]
\centerline{\psfig{file=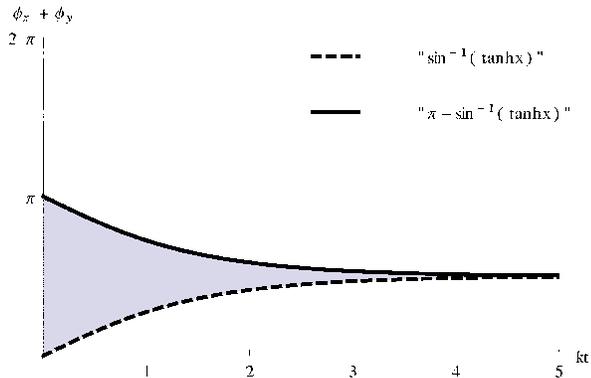,width=8cm}}
\vspace*{8pt}
\caption{Region of squeezing for $\phi_x+\phi_y$ with $kt$. Squeezing occurs outside the region between the curves}.\label{fig 1: Squeezing}
\end{figure}

\section{Result and Discussion}
The maximum degree of polarization squeezing is $\mathcal{D}_{1 max}=1-e^{-2T}$ for
$\theta=\pi/4$ and $(\phi_x+\phi_y)=3\pi/2$. This gives a very large amount of polarization 
squeezing at moderate interaction times $T$. For $T=1$, we get $\mathcal{D}_1=0.864$, {\it i.e.}, 
$86.4 \% $ polarization squeezing and for $T=2, \mathcal{D}_1=0.982$, {\it i.e.}, $ 98.2 \% $ 
polarization squeezing is obtained.

To adjust $\theta=\pi/4$ in experiments is easy as this requires only
the equal intensities of $x$ and $y$ components, {\it i.e.}, $|\alpha|=|\beta|$.
On this condition being followed alone, polarization squeezing is seen for
$\sqrt{R}=[\cosh2kt-\sinh2kt\sin(\phi_x+\phi_y)]>1$ or $ \sinh(\phi_x+\phi_y)<\tanh kt$. 

We plot $(\phi_x+\phi_y)$ against $kt$ [Figure \ref{fig 1: Squeezing}] and it shows that polarization squeezing occurs
for all values $(\phi_x+\phi_y)$ except those between $\phi_1=\sin^{-1}(\tanh kt)$ and 
$\phi_2=\pi-\sin^{-1}(\tanh kt)$, which is very narrow for appreciable values of interaction time.
For $\theta=\pi/4$, $\expec{\hat S_2(T)}$ and $\expec{\hat S_3(T)}$ take the form,
\begin{eqnarray}
\expec{\hat S_2(T)}&=& A^2\sinh2kt\cos(\phi_x-\phi_y), \nonumber\\
\expec{\hat S_3(T)}&=&-A^2\sin(\phi_x-\phi_y)e^{-2kt}.\nonumber
\end{eqnarray}
If we choose $\phi_x=\phi_y$, ${\hat S_2(T)}$ will have maximum value and ${\hat S_3(T)}$ will 
vanish. 
An experimentalist may therefore take plane polarized light along the direction dividing 
$x$ and $y$ directions ensuring $\phi_x=\phi_y$ and thus $\theta=\pi/4$ and then vary the input phase
$\phi_x=\phi_y$ so as to make 
\begin{enumerate}
\item ${\hat S_2(t)}=I_{+}-I_{-} $, where $I_{+}$ and $I_{-}$ are intensities of 
plane polarized components along directions midway between $x$ and $y$ directions.
\item ${\hat S_3(t)}=I_{R}-I_{L}=0$, where $I_{R}$ and $I_{L}$ are intensities of 
right handed and left handed circularly polarized components. 
\end{enumerate}
This will make $(\phi_x+\phi_y)=3\pi/2$ and the output will show maximum polarization squeezing.
\section*{Acknowledgement}

We would like to thank Hari Prakash and Naresh Chandra for their interest and critical comments.


\begin{thebibliography}{0}
\section{REFERENCES}
\bibitem{1}
G.~G.~Stokes, Trans.Cambridge Phylos. Soc.$\bm 9$, 399 (1852).

\bibitem{2} 
M.~Born and E.~Wolf, {\it Principles of Optics} (Cambridge Univ. Press, Cambridge, 1999).

\bibitem{3}
See, {\it e.g.}, ref.[2].

\bibitem{4}
L.~Mandel and E.~Wolf, {\it Optical Coherence and Quantum Optics} (Cambridge Univ. Press, 
Cambridge, 1995).

\bibitem{5} 
See, {\it e.g.}, ref.[4].

\bibitem{6} H.~Prakash and N.~Chandra, Phys. Rev. A ${\bm 4}$, 796 (1971); H.~Prakash and N.~Chandra, Phys. Rev. A 
$\bm 9$, 1021 (1974). 

\bibitem{7}
A.~S.~Chirkin and V.~V.~ Volokhovsky, Journal of Russian Laser Research ${\bm {16}}$, 6 (1995); 
N.~Korolkova and R.~Loudon, Phys. Rev. A ${\bm {71}}$, 032343 (2005).

\bibitem{8}
J.~Heersink, T.~Lorenz, O.~Glockl, N.~Korolkova,and G.~ Leuchs, Phys. Rev. A ${\bm {68}}$, 013815 (2003) .

\bibitem{9}
A.~S.~Chirkin, A.~A.~Orlov, and D.~Y.~Parashchuk, Quantum Electron ${\bm {23}}$, 870 (1993).

\bibitem{10}
A.~Luis and N.~Korolkova, Phys. Rev. A ${\bm {74}}$, 043817 (2006).

\bibitem{11}
R.~Prakash and N.~Shukla, Optics Communications ${\bm {284}}$, 3568 (2011).

\bibitem{12}
M.~O.~Schully and M.~S.~Zubairy, {\it Quantum Optics} (Cambridge Univ. Press, Cambridge, 1997).

\end{thebibliography}
\end{document}